\documentclass[3p,times,twocolumn]{elsarticle}
\pdfoutput=1

\usepackage{ecrc}
\usepackage{booktabs}
\usepackage{arydshln}
\usepackage{amsmath}

\volume{00}

\firstpage{1}

\journalname{Nuclear Physics B Proceedings Supplement}

\runauth{}

\jid{nuphbp}

\jnltitlelogo{Nuclear Physics B Proceedings Supplement}

\usepackage{amssymb}

\usepackage[figuresright]{rotating}

\def\ket#1{|{#1}\rangle}
\def\bra#1{\langle{#1}|}

\newcommand{\preprint}{DESY-14-183}

\begin{document}

\begin{frontmatter}



\dochead{}

\title{\vspace{-2em}{\small\hbox{}\hfill\preprint}\\[2em]GoSam 2.0: Automated one loop calculations within and beyond the Standard Model}


\author[mpi,desy]{Nicolas Greiner}

\address[mpi]{Max-Planck-Institut f\"ur Physik, F\"ohringer Ring 6, 80805 M\"unchen, Germany}
\address[desy]{DESY Theory Group, Notkestr. 85, D-22607 Hamburg, Germany}

\begin{abstract}
We present GoSam 2.0, a fully automated framework for the generation and evaluation of one loop amplitudes in multi leg
processes. The new version offers numerous improvements both on generational aspects as well as on the reduction side.
This leads to a faster and more stable code for calculations within and beyond the Standard Model. Furthermore it contains
the extended version of the standardized interface to Monte Carlo programs which allows for an easy combination with
other existing tools. We briefly describe
the conceptual innovations and present some phenomenological results.
\end{abstract}


\end{frontmatter}


\section{Introduction}
\label{intro}
Two of the main challenges for the upcoming run 2 of the LHC will be a more precise determination of the Higgs 
\cite{Aad:2012tfa,Chatrchyan:2012ufa} properties and
its couplings to bosons and fermions as well as the continued searches for new physics.\\
Both cases require a precise prediction for both signal and background processes. This particularly includes
the calculation of next-to-leading order corrections in QCD. One of the main bottlenecks of such a computation is the calculation
of the virtual one loop amplitude. The complexity and the need for having reliable tools for a large variety of different
processes has lead to the development of multi-purpose automated tools. An example of such a tool is the GoSam
package \cite{Cullen:2011ac} that focuses on the efficient generation and numerical evaluation of one loop amplitudes.
The continious refinement and extension of the existing package has lead to the publication of the version 2.0 \cite{Cullen:2014yla}.
In this talk we will describe the improvements and new features contained in the the new version and present
selected results that have been obtained with GoSam 2.0.

\section{New features in GoSam 2.0}
\subsection{Improvements in code generation}
\subsubsection{Code optimisation with FORM}
GoSam generates an algebraic expression for each amplitude which is written in a Fortran90 file. It is obvious
that the time needed to evaluate a single phase space point is highly dependent on how optimised the expression
is written. In the first version the generation of an optimised expression has been done with the help of haggies~\cite{Reiter:2009ts}.
In the new version we make use of new features provided by FORM version 4.x \cite{Kuipers:2012rf}.
The new features result in a more compact code and a gain in speed of up to an order of magnitude.
\subsubsection{Summing of diagrams with common subdiagrams}
In order to improve the efficiency and the evaluation time, GoSam 2.0 is able to automatically sum algebraically diagrams that exhibit
a similar structure to a 'meta-diagram', which  is then treated as a single diagram. In particular,
diagrams that differ only by a propagator, which is not in the loop (e.g. $Z$ vs. $\gamma$), are summed.
Also diagrams with the same loop, but with a different external tree part are summed up.
In the same way, diagrams, that share the same set of loop propagators but with different particle content in the loop,
are combined to a single diagram.\\
This summing is controlled by the option \texttt{diagsum}, which is set to \texttt{True} by default.
\subsubsection{Numerical polarisation vectors}
To reduce the size of the code, numerical polarisation vectors are used for massless gauge bosons. This means that
an algebraic expression is only written for a minimal set of helicity combinations and not for each combination separately.
Per default this option is used, however it can be switched off by setting \texttt{polvec=explicit}.
\subsection{Improvements in the reduction}
\subsubsection{New reduction method}
The default reduction method in GoSam 2.0 is \texttt{NINJA}~ \cite{Mastrolia:2012bu,vanDeurzen:2013saa,Peraro:2014cba}.
It is a further improvement of the integrand reduction method ~\cite{Ossola:2006us,Mastrolia:2008jb,Ossola:2008xq},
 based on the idea, that the coefficients of the residues
of a loop integral can be extracted in a more efficient way by performing a Laurent expansion of the integrand. This
methods requires less numerical sampling and therefore leads to a faster and more stable reduction.
\subsubsection{Higher rank integrals}
The tensor integral of a one loop calculation can be written in a very general form as
\begin{equation}
 I_{N}^{n,\mu_1,...\mu_r}(S)=\int d^nk\frac{k^{\mu_1}...k^{\mu_r}}{\prod^N_{i=1}\left((k+r_i)^2-m_i^2+i\delta\right)}
\end{equation}
In the Standard Model the maximal value for $r$ is $r=N$. However, in BSM theories and  effective theories, larger
values for $r$ can occur. Therefore, the libraries \texttt{NINJA}~\cite{Mastrolia:2012bu,vanDeurzen:2013saa,Peraro:2014cba},
\texttt{GOLEM95}~\cite{Binoth:2008uq,Heinrich:2010ax,Cullen:2011kv,Guillet:2013msa} and \texttt{SAMURAI}~\cite{Mastrolia:2010nb}
have been extended to deal with the case of $r=N+1$. This is an important ingredient for Higgs production in gluon fusion
which we discuss later.
\subsubsection{The \texttt{derive} extension}
The new version contains an improved tensorial reconstruction, based on the idea that the numerator can be Taylor expanded
around $q=0$,
\begin{flalign}
\mathcal{N}(q)&=\mathcal{N}(0)\nonumber \\
+ & q^\mu
  \frac{\partial}{\partial q_\mu}\mathcal{N}(q)\vert_{q=0}\nonumber \\
+& \frac1{2!} q^\mu q^\nu
  \frac{\partial}{\partial q_\mu}
  \frac{\partial}{\partial q_\nu}
  \mathcal{N}(q)\vert_{q=0} + \ldots \;.
\end{flalign}
This allows to read off the coefficients of the tensor integrals. It leads to a further improvement of the speed
and the precision of the tensorial reconstruction.
\subsubsection{Electroweak scheme choices}
There are a various number of electroweak schemes, depending on which parameters are used as input parameters and which parameters
are derived from the input paramaters. A consistent treatment requires that a minimal set of input parameters are given, and all
other parameters are then derived. GoSam 2.0 allows to choose all possible sets of consistent schemes and the remaining parameters
are then automatically derived.
\subsubsection{Rescue system}
The new release contains a rescue system to automatically detect and rescue numerically unstable points. Unstable points are
triggered by an insufficient cancellation of infrared poles. Several checks and re-evaluation with different reduction
methods can then be performed. For further details, see Refs.~\cite{Cullen:2014yla,vanDeurzen:2013saa}.

\subsection{New ranges of applicability}
\subsubsection{Color- and spin-correlated matrix elements}
The use of subtraction methods for NLO calculations require the calculation of color- and spin-correlated matrix elements,
i.e. born-like matrix elements with either a modified color structure or the combination of amplitudes where the helicity 
of one external leg is flipped. The color correlated matrix elements are defined as
\begin{equation}
 C_{ij}=\bra{{\cal M}}\textbf{T}_{i}\textbf{T}_j \ket{{\cal M}}\;,
\end{equation}
and the spin-correlated matrix element are defined as
\begin{equation}
 S_{ij}=\bra{{\cal M},-}\textbf{T}_{i}\textbf{T}_j \ket{{\cal M},+}\;.
\end{equation}
Bot color- and spin-correlated matrix elements contain implicitly
the sum over all helicities, only the helicities with the indices $i$ and $j$ are fixed, i.e.
 \begin{eqnarray}
\langle {\cal M}_{i,-} |{\mathbf T}_i\cdot {\mathbf T}_j |{\cal M}_{i,+}\rangle =\hspace{4cm}& \\ \nonumber
\sum_{\lambda_1,...,\lambda_{i-1},\lambda_{i+1},...,\lambda_n}
\langle {\cal M}_{\lambda_1,...,\lambda_{i-1},-,\lambda_{i+1},...,\lambda_n} |
{\mathbf T}_i\cdot {\mathbf T}_j | 
{\cal M}_{\lambda_1,...,\lambda_{i-1},+,\lambda_{i+1},...,\lambda_n}\rangle \;.
\end{eqnarray}
In the new release, GoSam is able to generate these matrix elements and the information
can be passed via the BLHA2 interface~\cite{Alioli:2013nda}.
\subsubsection{Complex mass scheme}
A gauge invariant treatment of massive gauge bosons requires the use of the complex mass scheme, where the
widths also enter in the definition of the weak mixing angle. The masses of the bosons are given by
\begin{equation}
 m_{V}^2 \to \mu_{V}^2 = m_{V}^2 -i m_{V} \Gamma_{V},\quad V=W,Z\;.
\end{equation}
In order to maintain gauge invariance this affects the definition of the Weinberg angle:
\begin{equation}
 \cos\theta_W^2 = \frac{\mu_W^2}{\mu_Z^2}\;.
\end{equation}
The complex mass scheme is implemented in GoSam via new models called \texttt{sm\_complex} and
\texttt{smdiag\_complex}, depending on whether one wants to use the full CKM matrix or a unit matrix.
\section{Phenomenological applications}
The new version GoSam 2.0 has been recently used in a sizeable number of challenging calculations of both
within and beyond the Standard Model~\cite{Cullen:2012eh,vanDeurzen:2013rv,Gehrmann:2013aga,Cullen:2013saa,Greiner:2013gca,
Gehrmann:2013bga,Dolan:2013rja,Luisoni:2013cuh,Hoeche:2013mua,vanDeurzen:2013xla,Heinrich:2013qaa,Butterworth:2014efa}.
In this talk we discuss the calculation of Higgs plus jets in gluon fusion and the calculation of a neutralino pair
 in association with a jet in the MSSM.
\subsection{Higgs plus jets in gluon fusion}
The gluon fusion channel is the dominant production mechanism of a Standard Model Higgs at the LHC. Even if one is interested
in the vector boson fusion channel the gluon fusion mechanism is an irreducible background and therefore its precise determination
is mandatory. In particular we have calculated the NLO QCD corrections to $H+2$~\cite{vanDeurzen:2013rv} and $H+3$~\cite{Cullen:2013saa},
and a comparison between the two processes has been studied in Ref.~\cite{Butterworth:2014efa}.
For the $H+2$ jets process the results have been obtained by interfacing GoSam with Sherpa~\cite{Gleisberg:2008ta} via the BLHA
interface~\cite{Binoth:2010xt}. In the case of $H+3$ jets we have used MadGraph~\cite{Stelzer:1994ta,Alwall:2007st} for the real
emission matrix element, and MadDipole~\cite{Frederix:2008hu,Frederix:2010cj} for the 
generation of the dipoles and the integrated subtraction terms. The phase space integration for these pieces has been performed using 
MadEvent~\cite{Maltoni:2002qb}, for the tree-level contribution and the integration of the virtual amplitude we have again
used Sherpa.\\
We have obtained the numerical results with a basic setup of 8 TeV center of mass energy, basic cuts on the jets with
$p_T >30$ GeV, $\eta <4.4$ and an anti-kt jet algorithm~\cite{Cacciari:2008gp,Cacciari:2011ma} of $R=0.4$. 
Renormalization- and factorization scales have been chosen to be equal and set to
$\mu_{F}=\mu_{R}=\frac{\hat{H}_T}{2}=\frac{1}{2}\left(\sqrt{m_{H}^{2}+p_{T,H}^{2}}+\sum_{i}|p_{T,i}|\right)$. For the LO PDFs
we have used the cteq6l1 pdf set, for the NLO PDFs we have used the ct10nlo pdf set.\\
The main results on the level of total cross sections are summarized in Table~\ref{hj23}. Both processes show a sizeable
global K-factor of roughly 1.3 which stresses the importance of including the NLO QCD corrections. The K-factor increases
if NLO PDFs are used for both LO and NLO calculation. One interesting aspect is the fact that if one looks at the ratios
of the total cross sections of H+3 over H+2 they show, they are to a very good approximation constant, independent whether one
looks at the ratio of LO cross sections with LO or NLO pdfs or at NLO cross sections.

\begin{table*}[t!]
  \centering\small
  \begin{tabular}{cp{10mm}lrlp{10mm}c}
    \toprule\\[-8pt]
    Sample     &&\multicolumn{3}{l}{Cross sections for Higgs boson plus}\\[4pt]
    $K$-factor && $\ge2$ jets & $f_3$ & $\ge3$ jets
    && $r_{3/2}$\\[4pt]
    \midrule\\[-10pt]
    & LO &\\[0pt] 
    \midrule\\[-7pt]
    $H$+2-jet2  \scriptsize(LO PDFs) && $1.23~^{+37\%}_{-24\%}$ \\[8pt]
    $H$+3-jets \scriptsize(LO PDFs) && $(0.381)$ & $1.0$ & $0.381~^{+53\%}_{-32\%}$
                                    && $0.310~^{0.347}_{0.278}$ \\[8pt]
    \hdashline\\[-5pt] 
    $H$+2-jets  \scriptsize(NLO PDFs)&& $0.970~^{+33\%}_{-23\%}$ \\[8pt]
    $H$+3-jets \scriptsize(NLO PDFs)&& $(0.286)$ & $1.0$ & $0.286~^{+50\%}_{-31\%}$
                                    && $0.295~^{0.332}_{0.265}$ \\[5pt]
    \midrule\\[-10pt]
    & NLO &\\[0pt] 
    \midrule\\[-7pt]
    $H$+2-jets   && $1.590~^{-4\%}_{-7\%}$ & $0.182$ &
                   $0.289~^{+49\%}_{-31\%}$ \\[8pt]
    $H$+3-jets  && $(0.485)$ & $1.0$ &
                   $0.485~^{-3\%}_{-13\%}$ &&
                   $0.305~^{0.307}_{0.284}$ \\[20pt]
    $K_2$, $K_3$ \scriptsize(LO PDFs for LO)
                && $1.29~~~^{0.911}_{1.59}$ && $1.27~~~^{0.806}_{1.63}$ \\[5pt]
    $K_2$, $K_3$ \scriptsize(NLO PDFs for LO)
                && $1.64~~~^{1.19}_{1.98}$ && $1.70~~~^{1.10}_{2.13}$ \\[5pt]
    \bottomrule
  \end{tabular}
  \caption{\label{hj23}
    Cross sections in pb for the various parton-level Higgs boson plus
    jet samples used in this study. The upper and lower parts of the
    table show the LO and NLO results, respectively, together with
    their uncertainties (in percent) from varying scales by factors of
    two, up (subscript position) and down (superscript position).
    NLO-to-LO $K$-factors, $K_n$, for both the inclusive 2-jets
    ($n=2$) and 3-jets ($n=3$) bin, the cross section ratio $r_{3/2}$
    and $m$-jet fractions, $f_m$, are given in addition.}
\end{table*}

The comparison between the two processes allows to asses the effects of the additional jet on Higgs observables.
\begin{figure}[t!]
 \centering
\includegraphics[width=0.5\textwidth]{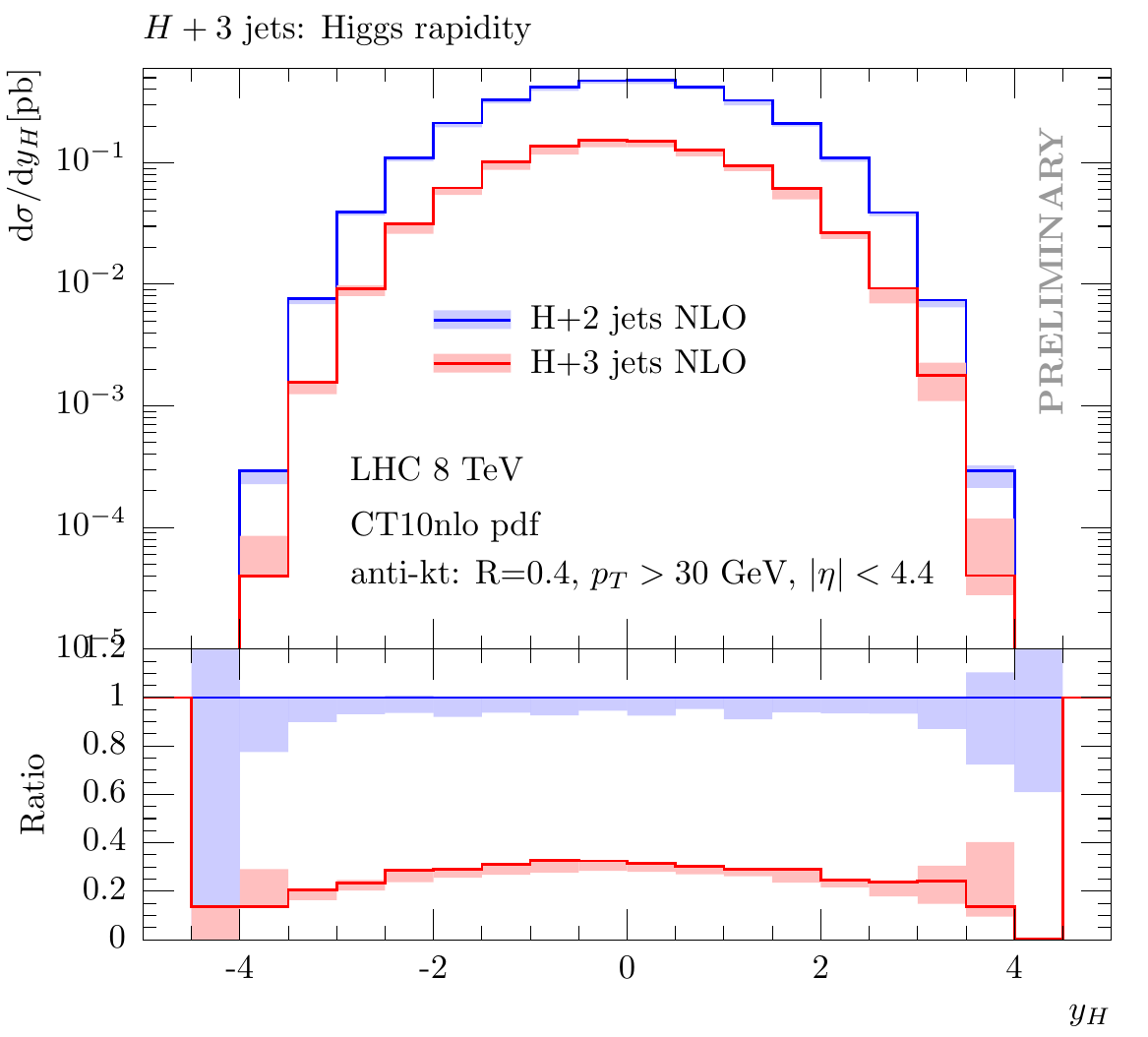}
\includegraphics[width=0.5\textwidth]{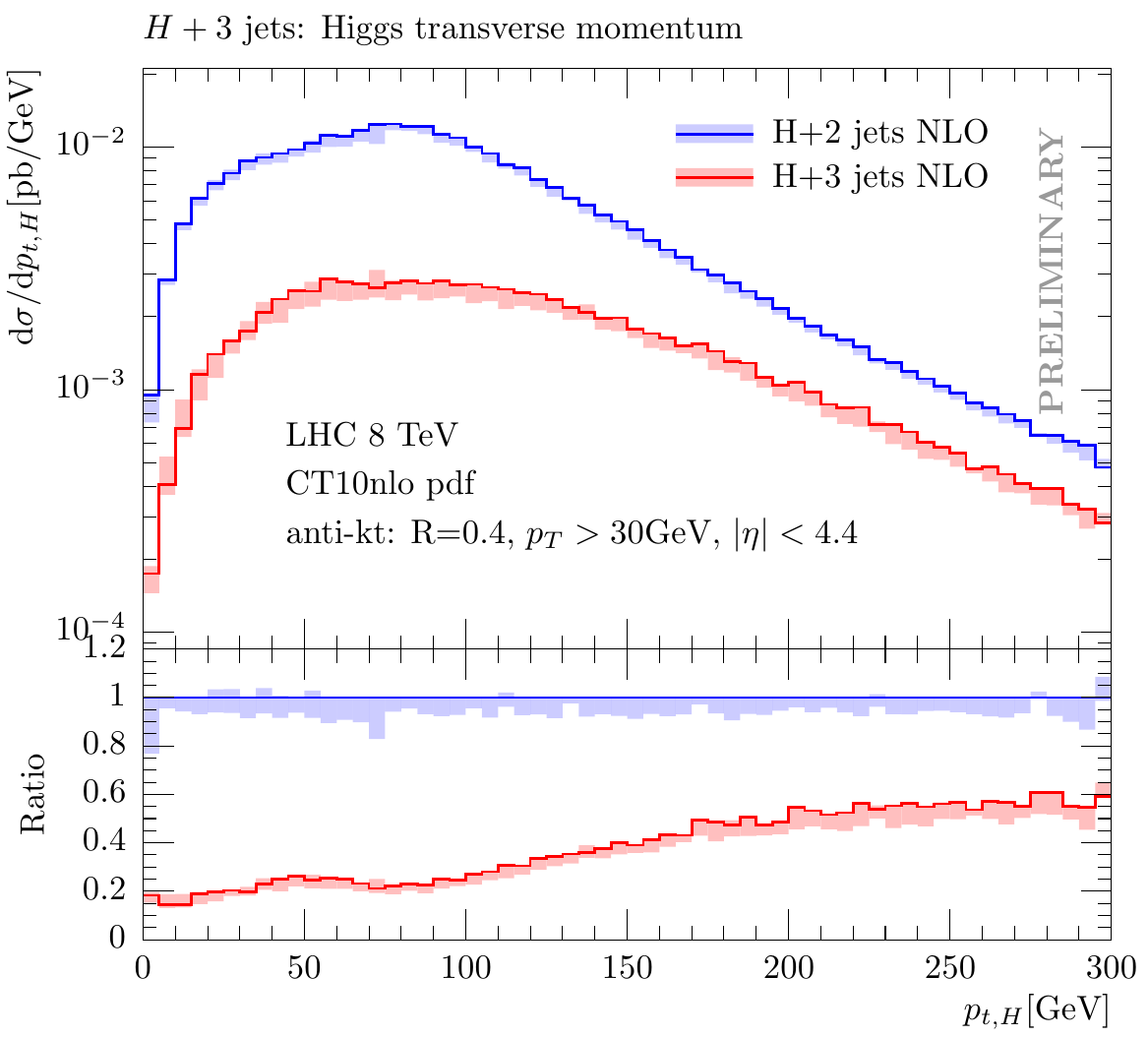}
\caption{Rapidity and $p_T$-distribution of the Higgs for $H+2$ and $H+3$ at NLO. The ratios are normalized to the $H+2$ result.}
\label{higgs:ypt}
\end{figure}
Two examples are shown in Fig.\ref{higgs:ypt}, namely the rapidity and the $p_T$ distribution of the Higgs. Looking at
the ratios allows us to asses the effect of the third jet on the observables. The ratio plots are normalized to the
$H+2$ result. One can see that the rapidity distribution is rather insensitive to the radiation of an additional jet, whereas
the $p_T$ distribution shows a clear increase of the importance of the third jet on the Higgs $p_T$ in the high-$p_T$ region.
 The reason for
this increase is a mere phase space argument. A high $p_T$ for the Higgs can be more easily obtained by distributing the necessary
recoil $p_T$ on three jets rather than on two. The high $p_T$ phase space points are almost evenly distributed in rapidity space,
therefore there is hardly any effect visible. 
\subsection{Neutralino pair production in association with a jet}
An example for a highly non-trivial BSM process is the production of a pair of the lightest neutralinos in 
association with a jet~\cite{Cullen:2012eh}. We have calculated the Susy-QCD corrections to this process in the MSSM~\cite{Cullen:2012eh}.
The neutralino is the LSP which makes this process lead to the simple experimental signature of missing energy and a mono-jet.
From a calculational point of view it is a very challenging process as it contains several mass scales. And as full 
off-shell effects are taken into account one has to deal with a non-trivial resonance structure.
The most complicated loop diagrams involve rank 3 pentagons 4 internal masses.
 Concerning the computational
setup GoSam has been used for the generation of the virtual one loop amplitude. In order to deal with a certain model (SM, MSSM, etc.) GoSam
requires the presence of a model file, that contains the Feynman rules of that model. Per default GoSam only contains model files
for various versions of the Standard Model, however new models can easily be imported with the help of Feynrules~\cite{Christensen:2008py}
which can be used to write a model file in the UFO format~\cite{Degrande:2011ua}. A model file in this format
is automatically understood by GoSam. Four our studies we chose a pragmatic and experimentally motivated 
 parameterisation of Susy, known as the
phenomenological MSSM (pMSSM)~\cite{Djouadi:2002ze,Berger:2008cq,AbdusSalam:2011fc}, in a variant involving 19 free parameters (p19MSSM).
The relevant Susy parameters are given in Table~\ref{SUSYparameters}.
\begin{table}
\begin{center}
{\small
\begin{tabular}{|l|l|}
\hline
\multicolumn{2}{|c|}{SUSY Parameters}\\
\hline
$M_{\tilde{\chi}_1^0} = 299.5$ & $\Gamma_ {\tilde{\chi}_1^0}= 0 $ \\
$M_{\tilde{g}}    = 415.9$ & $\Gamma_ {\tilde{g}}= 4.801$ \\
$M_{\tilde{u}_L} = 339.8$ & $\Gamma_ {\tilde{u}_L}= 0.002562$ \\
$M_{\tilde{u}_R} = 396.1$ & $\Gamma_ {\tilde{u}_R}= 0.1696$ \\
$M_{\tilde{d}_L} = 348.3$ & $\Gamma_ {\tilde{d}_L}= 0.003556$ \\
$M_{\tilde{d}_R} = 392.5$ & $\Gamma_ {\tilde{d}_R}= 0.04004$ \\
$M_{\tilde{b}_L} = 2518.0$ & $\Gamma_ {\tilde{b}_L}= 158.1$ \\
$M_{\tilde{b}_R} = 2541.8$ & $\Gamma_ {\tilde{b}_R}= 161.0$ \\
$M_{\tilde{t}_L} = 2403.7$ & $\Gamma_ {\tilde{t}_L}= 148.5$ \\
$M_{\tilde{t}_R} = 2668.6$ & $\Gamma_ {\tilde{t}_R}= 182.9$   \\
\hline
\end{tabular}
 }
\end{center}
\caption{Masses and widths of the supersymmetric particles for the benchmark point used. The second
generation of squarks is degenerate with the first generation of squarks. All parameters are given in GeV.\label{SUSYparameters}}
\end{table}

The calculation of the UV counter terms has been done separately. Tree-level and real emission matrix elements have been calculated
using MadGraph, for the subtraction terms we have used MadDipole.\\
For processes involving unstable particles, the proper definition of the set of diagrams
contributing to the next-to-leading order corrections is not obvious.
There are problems of double counting as diagrams with additional real radiation
from the unstable particle
in the final state can, if it becomes resonant, also be regarded as part of a leading order process
with the decay already included in the narrow width approximation. More specifically, in the real emission contribution
there is the possibility of producing a squark pair, where the squarks decay into a quark and a neutralino.
 Close to the resonance, this contribution gets
quite large, and in fact should rather be counted as a leading order contribution to
squark pair production with subsequent squark decay, because here we are interested in the radiative corrections
to the final state of a monojet in association with a neutralino pair.\\
Therefore the calculation was carried out in two different ways.
In the first approach we take into
account all possible diagrams leading to the required final state consisting of two neutralinos and two
QCD partons. In particular this includes the possibility of having two on-shell squarks.
In the second approach we remove the diagrams with two squarks in the s-channel from the amplitude.
In general, the removal of diagrams leads to a violation of gauge invariance, however one can show that
gauge invariance ist still preserved for a large class of gauges~\cite{Cullen:2012eh} or found to lead to a
small effect only~\cite{Frixione:2008yi}.

\begin{figure}
\includegraphics[width=0.5\textwidth]{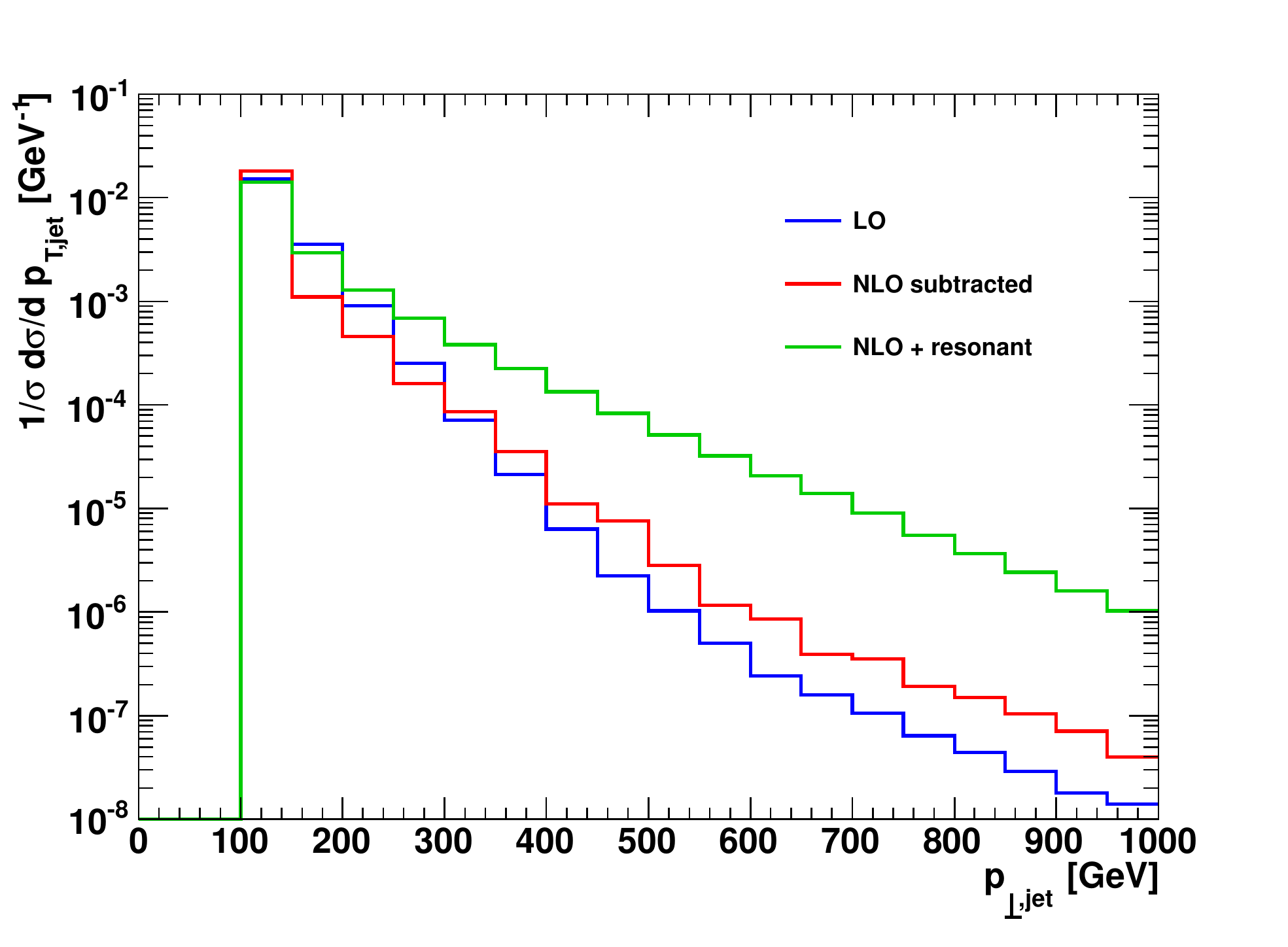} 
 \caption{Normalized distributions showing the transverse momentum distribution of the leading jet
at $\sqrt{s}=8$\,TeV, comparing the cases where the resonant diagrams are included to the ones where they are subtracted.
\label{fig:ptc} }
\end{figure}
The difference between the two approaches in case of the $p_T$ of the jet is shown in Fig.~\ref{fig:ptc}. 
The distribution is normalized to the total cross section. The curve in blue shows the distribution at leading order,
the red curve shows the NLO distribution where the doubly resonant squark pair diagrams have been removed. The green
curve shows the full result, also taking these resonant diagrams into account. As can be clearly seen these resonant
diagrams lead to a huge enhancement spoiling the perturbative convergence. Subtracting the diagrams then leads to a
well-behaved perturbative expansion. For a more detailed phenomenology of this process we refer to Ref.~\cite{Cullen:2012eh}.
 
\section{Conclusions}
In this talk we have presented the new release of GoSam, which contains a multitude of improvements compared to
the previous version. Refinements have been made in the context of diagram generation as well as on the reduction side,
both leading to a substantial gain in generation time, size of the code and the time needed for the evaluation of a 
phase space point. New reduction mechanisms and a rescue system have lead to a more stable and reliable performance.
We have discussed the new features and as selected examples for recent phenomenological applications we presented
the calculations of Higgs plus jets in gluon fusion and the production of a neutralino pair plus one jet in the
context of the MSSM.

\section{Acknowledgments}
We would like to thank the present and former members of the GoSam collaboration for their effort in the
development of GoSam. Furthermore we would like to thank Joey Huston, Jan Winter and Valery Yundin for their
collaboration and their work in the Higgs plus jets project.




\nocite{*}
\bibliographystyle{elsarticle-num}
\bibliography{refs}

\begin{thebibliography}{10}
\expandafter\ifx\csname url\endcsname\relax
  \def\url#1{\texttt{#1}}\fi
\expandafter\ifx\csname urlprefix\endcsname\relax\def\urlprefix{URL }\fi
\expandafter\ifx\csname href\endcsname\relax
  \def\href#1#2{#2} \def\path#1{#1}\fi

\bibitem{Aad:2012tfa}
G.~Aad, et~al., {Observation of a new particle in the search for the Standard
  Model Higgs boson with the ATLAS detector at the LHC}, Phys.Lett. B716 (2012)
  1--29.
\newblock \href {http://arxiv.org/abs/1207.7214} {\path{arXiv:1207.7214}},
  \href {http://dx.doi.org/10.1016/j.physletb.2012.08.020}
  {\path{doi:10.1016/j.physletb.2012.08.020}}.

\bibitem{Chatrchyan:2012ufa}
S.~Chatrchyan, et~al., {Observation of a new boson at a mass of 125 GeV with
  the CMS experiment at the LHC}, Phys.Lett. B716 (2012) 30--61.
\newblock \href {http://arxiv.org/abs/1207.7235} {\path{arXiv:1207.7235}},
  \href {http://dx.doi.org/10.1016/j.physletb.2012.08.021}
  {\path{doi:10.1016/j.physletb.2012.08.021}}.

\bibitem{Cullen:2011ac}
G.~Cullen, N.~Greiner, G.~Heinrich, G.~Luisoni, P.~Mastrolia, et~al.,
  {Automated One-Loop Calculations with GoSam}, Eur.Phys.J. C72 (2012) 1889.
\newblock \href {http://arxiv.org/abs/1111.2034} {\path{arXiv:1111.2034}},
  \href {http://dx.doi.org/10.1140/epjc/s10052-012-1889-1}
  {\path{doi:10.1140/epjc/s10052-012-1889-1}}.

\bibitem{Cullen:2014yla}
G.~Cullen, H.~van Deurzen, N.~Greiner, G.~Heinrich, G.~Luisoni, et~al.,
  {GoSam-2.0: a tool for automated one-loop calculations within the Standard
  Model and beyond}, Eur.Phys.J. C74 (2014) 3001.
\newblock \href {http://arxiv.org/abs/1404.7096} {\path{arXiv:1404.7096}},
  \href {http://dx.doi.org/10.1140/epjc/s10052-014-3001-5}
  {\path{doi:10.1140/epjc/s10052-014-3001-5}}.

\bibitem{Reiter:2009ts}
T.~Reiter, {Optimising Code Generation with haggies}, Comput.Phys.Commun. 181
  (2010) 1301--1331.
\newblock \href {http://arxiv.org/abs/0907.3714} {\path{arXiv:0907.3714}},
  \href {http://dx.doi.org/10.1016/j.cpc.2010.01.012}
  {\path{doi:10.1016/j.cpc.2010.01.012}}.

\bibitem{Kuipers:2012rf}
J.~Kuipers, T.~Ueda, J.~Vermaseren, J.~Vollinga, {FORM version 4.0},
  Comput.Phys.Commun. 184 (2013) 1453--1467.
\newblock \href {http://arxiv.org/abs/1203.6543} {\path{arXiv:1203.6543}},
  \href {http://dx.doi.org/10.1016/j.cpc.2012.12.028}
  {\path{doi:10.1016/j.cpc.2012.12.028}}.

\bibitem{Mastrolia:2012bu}
P.~Mastrolia, E.~Mirabella, T.~Peraro, {Integrand reduction of one-loop
  scattering amplitudes through Laurent series expansion}, JHEP 1206 (2012)
  095.
\newblock \href {http://arxiv.org/abs/1203.0291} {\path{arXiv:1203.0291}},
  \href {http://dx.doi.org/10.1007/JHEP11(2012)128, 10.1007/JHEP06(2012)095}
  {\path{doi:10.1007/JHEP11(2012)128, 10.1007/JHEP06(2012)095}}.

\bibitem{vanDeurzen:2013saa}
H.~van Deurzen, G.~Luisoni, P.~Mastrolia, E.~Mirabella, G.~Ossola, et~al.,
  {Multi-leg One-loop Massive Amplitudes from Integrand Reduction via Laurent
  Expansion}, JHEP 1403 (2014) 115.
\newblock \href {http://arxiv.org/abs/1312.6678} {\path{arXiv:1312.6678}},
  \href {http://dx.doi.org/10.1007/JHEP03(2014)115}
  {\path{doi:10.1007/JHEP03(2014)115}}.

\bibitem{Peraro:2014cba}
T.~Peraro, {Ninja: Automated Integrand Reduction via Laurent Expansion for
  One-Loop Amplitudes}, Comput.Phys.Commun. 185 (2014) 2771--2797.
\newblock \href {http://arxiv.org/abs/1403.1229} {\path{arXiv:1403.1229}},
  \href {http://dx.doi.org/10.1016/j.cpc.2014.06.017}
  {\path{doi:10.1016/j.cpc.2014.06.017}}.

\bibitem{Ossola:2006us}
G.~Ossola, C.~G. Papadopoulos, R.~Pittau, {Reducing full one-loop amplitudes to
  scalar integrals at the integrand level}, Nucl.Phys. B763 (2007) 147--169.
\newblock \href {http://arxiv.org/abs/hep-ph/0609007}
  {\path{arXiv:hep-ph/0609007}}, \href
  {http://dx.doi.org/10.1016/j.nuclphysb.2006.11.012}
  {\path{doi:10.1016/j.nuclphysb.2006.11.012}}.

\bibitem{Mastrolia:2008jb}
P.~Mastrolia, G.~Ossola, C.~Papadopoulos, R.~Pittau, {Optimizing the Reduction
  of One-Loop Amplitudes}, JHEP 0806 (2008) 030.
\newblock \href {http://arxiv.org/abs/0803.3964} {\path{arXiv:0803.3964}},
  \href {http://dx.doi.org/10.1088/1126-6708/2008/06/030}
  {\path{doi:10.1088/1126-6708/2008/06/030}}.

\bibitem{Ossola:2008xq}
G.~Ossola, C.~G. Papadopoulos, R.~Pittau, {On the Rational Terms of the
  one-loop amplitudes}, JHEP 0805 (2008) 004.
\newblock \href {http://arxiv.org/abs/0802.1876} {\path{arXiv:0802.1876}},
  \href {http://dx.doi.org/10.1088/1126-6708/2008/05/004}
  {\path{doi:10.1088/1126-6708/2008/05/004}}.

\bibitem{Binoth:2008uq}
T.~Binoth, J.-P. Guillet, G.~Heinrich, E.~Pilon, T.~Reiter, {Golem95: A
  Numerical program to calculate one-loop tensor integrals with up to six
  external legs}, Comput.Phys.Commun. 180 (2009) 2317--2330.
\newblock \href {http://arxiv.org/abs/0810.0992} {\path{arXiv:0810.0992}},
  \href {http://dx.doi.org/10.1016/j.cpc.2009.06.024}
  {\path{doi:10.1016/j.cpc.2009.06.024}}.

\bibitem{Heinrich:2010ax}
G.~Heinrich, G.~Ossola, T.~Reiter, F.~Tramontano, {Tensorial Reconstruction at
  the Integrand Level}, JHEP 1010 (2010) 105.
\newblock \href {http://arxiv.org/abs/1008.2441} {\path{arXiv:1008.2441}},
  \href {http://dx.doi.org/10.1007/JHEP10(2010)105}
  {\path{doi:10.1007/JHEP10(2010)105}}.

\bibitem{Cullen:2011kv}
G.~Cullen, J.~P. Guillet, G.~Heinrich, T.~Kleinschmidt, E.~Pilon, et~al.,
  {Golem95C: A library for one-loop integrals with complex masses},
  Comput.Phys.Commun. 182 (2011) 2276--2284.
\newblock \href {http://arxiv.org/abs/1101.5595} {\path{arXiv:1101.5595}},
  \href {http://dx.doi.org/10.1016/j.cpc.2011.05.015}
  {\path{doi:10.1016/j.cpc.2011.05.015}}.

\bibitem{Guillet:2013msa}
J.~P. Guillet, G.~Heinrich, J.~von Soden-Fraunhofen, {Tools for NLO automation:
  extension of the golem95C integral library}, Comput.Phys.Commun. 185 (2014)
  1828--1834.
\newblock \href {http://arxiv.org/abs/1312.3887} {\path{arXiv:1312.3887}},
  \href {http://dx.doi.org/10.1016/j.cpc.2014.03.009}
  {\path{doi:10.1016/j.cpc.2014.03.009}}.

\bibitem{Mastrolia:2010nb}
P.~Mastrolia, G.~Ossola, T.~Reiter, F.~Tramontano, {Scattering AMplitudes from
  Unitarity-based Reduction Algorithm at the Integrand-level}, JHEP 1008 (2010)
  080.
\newblock \href {http://arxiv.org/abs/1006.0710} {\path{arXiv:1006.0710}},
  \href {http://dx.doi.org/10.1007/JHEP08(2010)080}
  {\path{doi:10.1007/JHEP08(2010)080}}.

\bibitem{Alioli:2013nda}
S.~Alioli, S.~Badger, J.~Bellm, B.~Biedermann, F.~Boudjema, et~al., {Update of
  the Binoth Les Houches Accord for a standard interface between Monte Carlo
  tools and one-loop programs}, Comput.Phys.Commun. 185 (2014) 560--571.
\newblock \href {http://arxiv.org/abs/1308.3462} {\path{arXiv:1308.3462}},
  \href {http://dx.doi.org/10.1016/j.cpc.2013.10.020}
  {\path{doi:10.1016/j.cpc.2013.10.020}}.

\bibitem{Cullen:2012eh}
G.~Cullen, N.~Greiner, G.~Heinrich, {Susy-QCD corrections to neutralino pair
  production in association with a jet}, Eur.Phys.J. C73 (2013) 2388.
\newblock \href {http://arxiv.org/abs/1212.5154} {\path{arXiv:1212.5154}},
  \href {http://dx.doi.org/10.1140/epjc/s10052-013-2388-8}
  {\path{doi:10.1140/epjc/s10052-013-2388-8}}.

\bibitem{vanDeurzen:2013rv}
H.~van Deurzen, N.~Greiner, G.~Luisoni, P.~Mastrolia, E.~Mirabella, et~al.,
  {NLO QCD corrections to the production of Higgs plus two jets at the LHC},
  Phys.Lett. B721 (2013) 74--81.
\newblock \href {http://arxiv.org/abs/1301.0493} {\path{arXiv:1301.0493}},
  \href {http://dx.doi.org/10.1016/j.physletb.2013.02.051}
  {\path{doi:10.1016/j.physletb.2013.02.051}}.

\bibitem{Gehrmann:2013aga}
T.~Gehrmann, N.~Greiner, G.~Heinrich, {Photon isolation effects at NLO in
  $\gamma \gamma$ + jet final states in hadronic collisions}, JHEP 1306 (2013)
  058.
\newblock \href {http://arxiv.org/abs/1303.0824} {\path{arXiv:1303.0824}},
  \href {http://dx.doi.org/10.1007/JHEP06(2014)076, 10.1007/JHEP06(2013)058}
  {\path{doi:10.1007/JHEP06(2014)076, 10.1007/JHEP06(2013)058}}.

\bibitem{Cullen:2013saa}
G.~Cullen, H.~van Deurzen, N.~Greiner, G.~Luisoni, P.~Mastrolia, et~al.,
  {Next-to-Leading-Order QCD Corrections to Higgs Boson Production Plus Three
  Jets in Gluon Fusion}, Phys.Rev.Lett. 111~(13) (2013) 131801.
\newblock \href {http://arxiv.org/abs/1307.4737} {\path{arXiv:1307.4737}},
  \href {http://dx.doi.org/10.1103/PhysRevLett.111.131801}
  {\path{doi:10.1103/PhysRevLett.111.131801}}.

\bibitem{Greiner:2013gca}
N.~Greiner, G.~Heinrich, J.~Reichel, J.~F. von Soden-Fraunhofen, {NLO QCD
  Corrections to Diphoton Plus Jet Production through Graviton Exchange}, JHEP
  1311 (2013) 028.
\newblock \href {http://arxiv.org/abs/1308.2194} {\path{arXiv:1308.2194}},
  \href {http://dx.doi.org/10.1007/JHEP11(2013)028}
  {\path{doi:10.1007/JHEP11(2013)028}}.

\bibitem{Gehrmann:2013bga}
T.~Gehrmann, N.~Greiner, G.~Heinrich, {Precise QCD predictions for the
  production of a photon pair in association with two jets}, Phys.Rev.Lett. 111
  (2013) 222002.
\newblock \href {http://arxiv.org/abs/1308.3660} {\path{arXiv:1308.3660}},
  \href {http://dx.doi.org/10.1103/PhysRevLett.111.222002}
  {\path{doi:10.1103/PhysRevLett.111.222002}}.

\bibitem{Dolan:2013rja}
M.~J. Dolan, C.~Englert, N.~Greiner, M.~Spannowsky, {Further on up the road:
  $hhjj$ production at the LHC}, Phys.Rev.Lett. 112 (2014) 101802.
\newblock \href {http://arxiv.org/abs/1310.1084} {\path{arXiv:1310.1084}},
  \href {http://dx.doi.org/10.1103/PhysRevLett.112.101802}
  {\path{doi:10.1103/PhysRevLett.112.101802}}.

\bibitem{Luisoni:2013cuh}
G.~Luisoni, P.~Nason, C.~Oleari, F.~Tramontano, {$HW^{\pm}$/HZ + 0 and 1 jet at
  NLO with the POWHEG BOX interfaced to GoSam and their merging within MiNLO},
  JHEP 1310 (2013) 083.
\newblock \href {http://arxiv.org/abs/1306.2542} {\path{arXiv:1306.2542}},
  \href {http://dx.doi.org/10.1007/JHEP10(2013)083}
  {\path{doi:10.1007/JHEP10(2013)083}}.

\bibitem{Hoeche:2013mua}
S.~Hoeche, J.~Huang, G.~Luisoni, M.~Schoenherr, J.~Winter, {Zero and one jet
  combined next-to-leading order analysis of the top quark forward-backward
  asymmetry}, Phys.Rev. D88~(1) (2013) 014040.
\newblock \href {http://arxiv.org/abs/1306.2703} {\path{arXiv:1306.2703}},
  \href {http://dx.doi.org/10.1103/PhysRevD.88.014040}
  {\path{doi:10.1103/PhysRevD.88.014040}}.

\bibitem{vanDeurzen:2013xla}
H.~van Deurzen, G.~Luisoni, P.~Mastrolia, E.~Mirabella, G.~Ossola, et~al.,
  {Next-to-Leading-Order QCD Corrections to Higgs Boson Production in
  Association with a Top Quark Pair and a Jet}, Phys.Rev.Lett. 111~(17) (2013)
  171801.
\newblock \href {http://arxiv.org/abs/1307.8437} {\path{arXiv:1307.8437}},
  \href {http://dx.doi.org/10.1103/PhysRevLett.111.171801}
  {\path{doi:10.1103/PhysRevLett.111.171801}}.

\bibitem{Heinrich:2013qaa}
G.~Heinrich, A.~Maier, R.~Nisius, J.~Schlenk, J.~Winter, {NLO QCD corrections
  to $W^{+} W^{-}b\bar{b}$ production with leptonic decays in the light of top
  quark mass and asymmetry measurements}, JHEP 1406 (2014) 158.
\newblock \href {http://arxiv.org/abs/1312.6659} {\path{arXiv:1312.6659}},
  \href {http://dx.doi.org/10.1007/JHEP06(2014)158}
  {\path{doi:10.1007/JHEP06(2014)158}}.

\bibitem{Butterworth:2014efa}
J.~Butterworth, G.~Dissertori, S.~Dittmaier, D.~de~Florian, N.~Glover, et~al.,
  {Les Houches 2013: Physics at TeV Colliders: Standard Model Working Group
  Report}\href {http://arxiv.org/abs/1405.1067} {\path{arXiv:1405.1067}}.

\bibitem{Gleisberg:2008ta}
T.~Gleisberg, S.~Hoeche, F.~Krauss, M.~Schonherr, S.~Schumann, et~al., {Event
  generation with SHERPA 1.1}, JHEP 0902 (2009) 007.
\newblock \href {http://arxiv.org/abs/0811.4622} {\path{arXiv:0811.4622}},
  \href {http://dx.doi.org/10.1088/1126-6708/2009/02/007}
  {\path{doi:10.1088/1126-6708/2009/02/007}}.

\bibitem{Binoth:2010xt}
T.~Binoth, F.~Boudjema, G.~Dissertori, A.~Lazopoulos, A.~Denner, et~al., {A
  Proposal for a standard interface between Monte Carlo tools and one-loop
  programs}, Comput.Phys.Commun. 181 (2010) 1612--1622.
\newblock \href {http://arxiv.org/abs/1001.1307} {\path{arXiv:1001.1307}},
  \href {http://dx.doi.org/10.1016/j.cpc.2010.05.016}
  {\path{doi:10.1016/j.cpc.2010.05.016}}.

\bibitem{Stelzer:1994ta}
T.~Stelzer, W.~Long, {Automatic generation of tree level helicity amplitudes},
  Comput.Phys.Commun. 81 (1994) 357--371.
\newblock \href {http://arxiv.org/abs/hep-ph/9401258}
  {\path{arXiv:hep-ph/9401258}}, \href
  {http://dx.doi.org/10.1016/0010-4655(94)90084-1}
  {\path{doi:10.1016/0010-4655(94)90084-1}}.

\bibitem{Alwall:2007st}
J.~Alwall, P.~Demin, S.~de~Visscher, R.~Frederix, M.~Herquet, et~al.,
  {MadGraph/MadEvent v4: The New Web Generation}, JHEP 0709 (2007) 028.
\newblock \href {http://arxiv.org/abs/0706.2334} {\path{arXiv:0706.2334}},
  \href {http://dx.doi.org/10.1088/1126-6708/2007/09/028}
  {\path{doi:10.1088/1126-6708/2007/09/028}}.

\bibitem{Frederix:2008hu}
R.~Frederix, T.~Gehrmann, N.~Greiner, {Automation of the Dipole Subtraction
  Method in MadGraph/MadEvent}, JHEP 0809 (2008) 122.
\newblock \href {http://arxiv.org/abs/0808.2128} {\path{arXiv:0808.2128}},
  \href {http://dx.doi.org/10.1088/1126-6708/2008/09/122}
  {\path{doi:10.1088/1126-6708/2008/09/122}}.

\bibitem{Frederix:2010cj}
R.~Frederix, T.~Gehrmann, N.~Greiner, {Integrated dipoles with MadDipole in the
  MadGraph framework}, JHEP 1006 (2010) 086.
\newblock \href {http://arxiv.org/abs/1004.2905} {\path{arXiv:1004.2905}},
  \href {http://dx.doi.org/10.1007/JHEP06(2010)086}
  {\path{doi:10.1007/JHEP06(2010)086}}.

\bibitem{Maltoni:2002qb}
F.~Maltoni, T.~Stelzer, {MadEvent: Automatic event generation with MadGraph},
  JHEP 0302 (2003) 027.
\newblock \href {http://arxiv.org/abs/hep-ph/0208156}
  {\path{arXiv:hep-ph/0208156}}, \href
  {http://dx.doi.org/10.1088/1126-6708/2003/02/027}
  {\path{doi:10.1088/1126-6708/2003/02/027}}.

\bibitem{Cacciari:2008gp}
M.~Cacciari, G.~P. Salam, G.~Soyez, {The Anti-k(t) jet clustering algorithm},
  JHEP 0804 (2008) 063.
\newblock \href {http://arxiv.org/abs/0802.1189} {\path{arXiv:0802.1189}},
  \href {http://dx.doi.org/10.1088/1126-6708/2008/04/063}
  {\path{doi:10.1088/1126-6708/2008/04/063}}.

\bibitem{Cacciari:2011ma}
M.~Cacciari, G.~P. Salam, G.~Soyez, {FastJet User Manual}, Eur.Phys.J. C72
  (2012) 1896.
\newblock \href {http://arxiv.org/abs/1111.6097} {\path{arXiv:1111.6097}},
  \href {http://dx.doi.org/10.1140/epjc/s10052-012-1896-2}
  {\path{doi:10.1140/epjc/s10052-012-1896-2}}.

\bibitem{Christensen:2008py}
N.~D. Christensen, C.~Duhr, {FeynRules - Feynman rules made easy},
  Comput.Phys.Commun. 180 (2009) 1614--1641.
\newblock \href {http://arxiv.org/abs/0806.4194} {\path{arXiv:0806.4194}},
  \href {http://dx.doi.org/10.1016/j.cpc.2009.02.018}
  {\path{doi:10.1016/j.cpc.2009.02.018}}.

\bibitem{Degrande:2011ua}
C.~Degrande, C.~Duhr, B.~Fuks, D.~Grellscheid, O.~Mattelaer, et~al., {UFO - The
  Universal FeynRules Output}, Comput.Phys.Commun. 183 (2012) 1201--1214.
\newblock \href {http://arxiv.org/abs/1108.2040} {\path{arXiv:1108.2040}},
  \href {http://dx.doi.org/10.1016/j.cpc.2012.01.022}
  {\path{doi:10.1016/j.cpc.2012.01.022}}.

\bibitem{Djouadi:2002ze}
A.~Djouadi, J.-L. Kneur, G.~Moultaka, {SuSpect: A Fortran code for the
  supersymmetric and Higgs particle spectrum in the MSSM}, Comput.Phys.Commun.
  176 (2007) 426--455.
\newblock \href {http://arxiv.org/abs/hep-ph/0211331}
  {\path{arXiv:hep-ph/0211331}}, \href
  {http://dx.doi.org/10.1016/j.cpc.2006.11.009}
  {\path{doi:10.1016/j.cpc.2006.11.009}}.

\bibitem{Berger:2008cq}
C.~F. Berger, J.~S. Gainer, J.~L. Hewett, T.~G. Rizzo, {Supersymmetry Without
  Prejudice}, JHEP 0902 (2009) 023.
\newblock \href {http://arxiv.org/abs/0812.0980} {\path{arXiv:0812.0980}},
  \href {http://dx.doi.org/10.1088/1126-6708/2009/02/023}
  {\path{doi:10.1088/1126-6708/2009/02/023}}.

\bibitem{AbdusSalam:2011fc}
S.~AbdusSalam, B.~Allanach, H.~Dreiner, J.~Ellis, U.~Ellwanger, et~al.,
  {Benchmark Models, Planes, Lines and Points for Future SUSY Searches at the
  LHC}, Eur.Phys.J. C71 (2011) 1835.
\newblock \href {http://arxiv.org/abs/1109.3859} {\path{arXiv:1109.3859}}.

\bibitem{Frixione:2008yi}
S.~Frixione, E.~Laenen, P.~Motylinski, B.~R. Webber, C.~D. White, {Single-top
  hadroproduction in association with a W boson}, JHEP 0807 (2008) 029.
\newblock \href {http://arxiv.org/abs/0805.3067} {\path{arXiv:0805.3067}},
  \href {http://dx.doi.org/10.1088/1126-6708/2008/07/029}
  {\path{doi:10.1088/1126-6708/2008/07/029}}.

\end{thebibliography}







\end{document}